\documentclass[%
 reprint,
superscriptaddress,
 amsmath,amssymb,
 aps,
 prl,
]{revtex4-1}
\usepackage{graphicx}
\usepackage{dcolumn}
\usepackage{bm}

\usepackage{overpic}
\usepackage{color}
\begin{document}

\title{Injection induced by coaxial laser interference in laser wakefield accelerators}

\author{Jia Wang}
\affiliation{Institute of High Energy Physics, Chinese Academy of Sciences, Beijing 100049, China}
\affiliation{University of Chinese Academy of Sciences, Beijing 100049, China}
\author{Ming Zeng}
 \email[Corresponding author: ]{zengming@ihep.ac.cn}
\affiliation{Institute of High Energy Physics, Chinese Academy of Sciences, Beijing 100049, China}
\affiliation{University of Chinese Academy of Sciences, Beijing 100049, China}
\author{Dazhang Li}
 \email[Corresponding author: ]{lidz@ihep.ac.cn}
\affiliation{Institute of High Energy Physics, Chinese Academy of Sciences, Beijing 100049, China}
\affiliation{University of Chinese Academy of Sciences, Beijing 100049, China}
\author{Xiaoning Wang}
\affiliation{Institute of High Energy Physics, Chinese Academy of Sciences, Beijing 100049, China}
\affiliation{University of Chinese Academy of Sciences, Beijing 100049, China}
\author{Wei Lu}
\affiliation{Department of Engineering Physics, Tsinghua University, Beijing 100084, China}
\author{Jie Gao}
\affiliation{Institute of High Energy Physics, Chinese Academy of Sciences, Beijing 100049, China}
\affiliation{University of Chinese Academy of Sciences, Beijing 100049, China}

\date{\today}

\begin{abstract}
A new injection scheme using the interference of two coaxial laser pulses is proposed for generating high quality beams in laser wakefield accelerators. In this scheme, a relatively loosely focused laser pulse drives the plasma wakefield, and a tightly focused laser pulse with similar intensity triggers interference ring pattern which creates onion-like multi sheaths in the plasma wakefield. Due to the wavefront curvature change after the focal position of the tightly focused laser, the innermost sheath of the wakefield expands, which slows down the effective phase velocity of the wakefield and triggers injection of plasma electrons. Particle-in-cell simulations show that high quality electron beams with low energy spread (a few per mill), high charge (hundred picocoulomb) and small emittance (sub millimeter milliradian) at the same time can be generated using moderate laser parameters for properly chosen phase differences between the two lasers.
\end{abstract}
\maketitle

Laser wakefield accelerator (LWFA) has undergone tremendous progresses in the past decades and is becoming one of the most important candidates for the next generation accelerator based light sources and colliders~\cite{TTajimaPRL1979, SPDManglesNature2004, JFaureNature2004, CGRGeddesNature2004, WTWangNature2021}. Currently, the energy record in LWFA experiments has reached 7.8 GeV~\cite{WPLeemansPRL2014, AJGonsalvesPRL2019}, and the typical energy spread is in a few percent range, with the lowest reported value slightly below 1\%, with the beam charges of 10s of pC~\cite{ADoppPRL2018, WTWangPRL2016, KeLPRL2021}. However, for challenging applications like plasma-based colliders and free-electron-lasers, orders of magnitude better beam qualities are required, and need to be achieved simultaneously, and this motivates a persistent searching for methods of generating high quality beams~\cite{LeeNP2006,PakPRL2010,ViePRL2011,ABuckPRL2013}.

Among these methods, all-optical injection schemes are very attractive due to their simple setups and potential for precise controllability~\cite{UmsPRL1996, EsaPRL1997, ZMShengEPJST2009}. Experimental demonstrations of these schemes and their modified versions typically produce electron beams with the energy spread $>1\%$ and the charges of 10s of pC~\cite{JFNature2006, ThomasPRL2008, KotPRL2009,RecPRL2009,GolPRL2018, QChenPRL2022}. Further optimization of these schemes based on particle-in-cell (PIC) simulations have suggested better beam qualities~\cite{DavPRL2009, LehePRL2013, HuPRAB2016, ZengNJP2020, WangPPCF2022}, however, none of these studies have produced a beam with simultaneously small energy spread of a few per mill, small emittance below half millimeter milliradian ($\rm mm\cdot mrad$), and large amount of charge over 100 pC.

\begin{figure}
    \centering
    \begin{overpic}[width=0.24\textwidth]{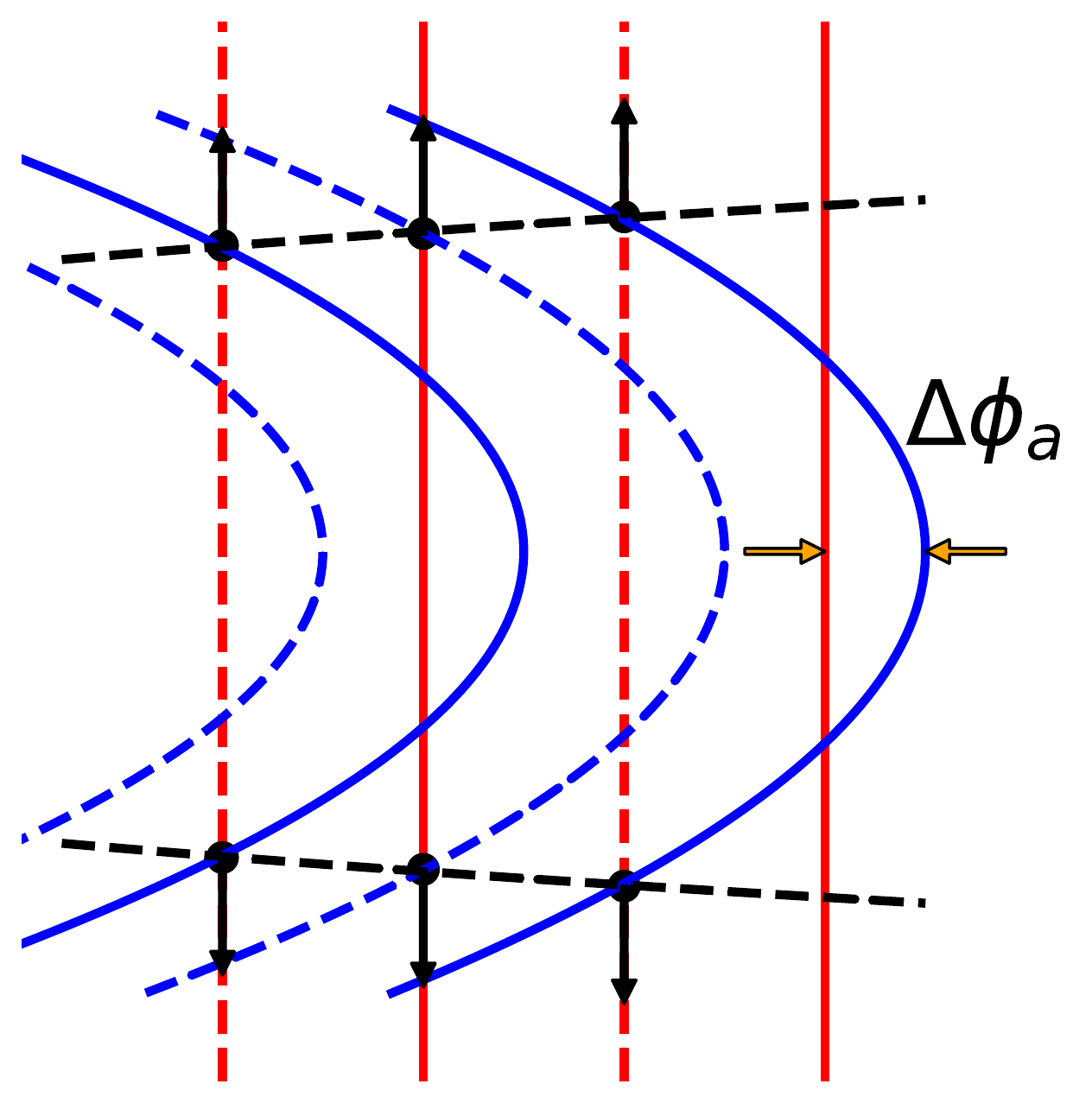}
        \put(82,5){(a)}
    \end{overpic}
    \begin{overpic}[width=0.22\textwidth]{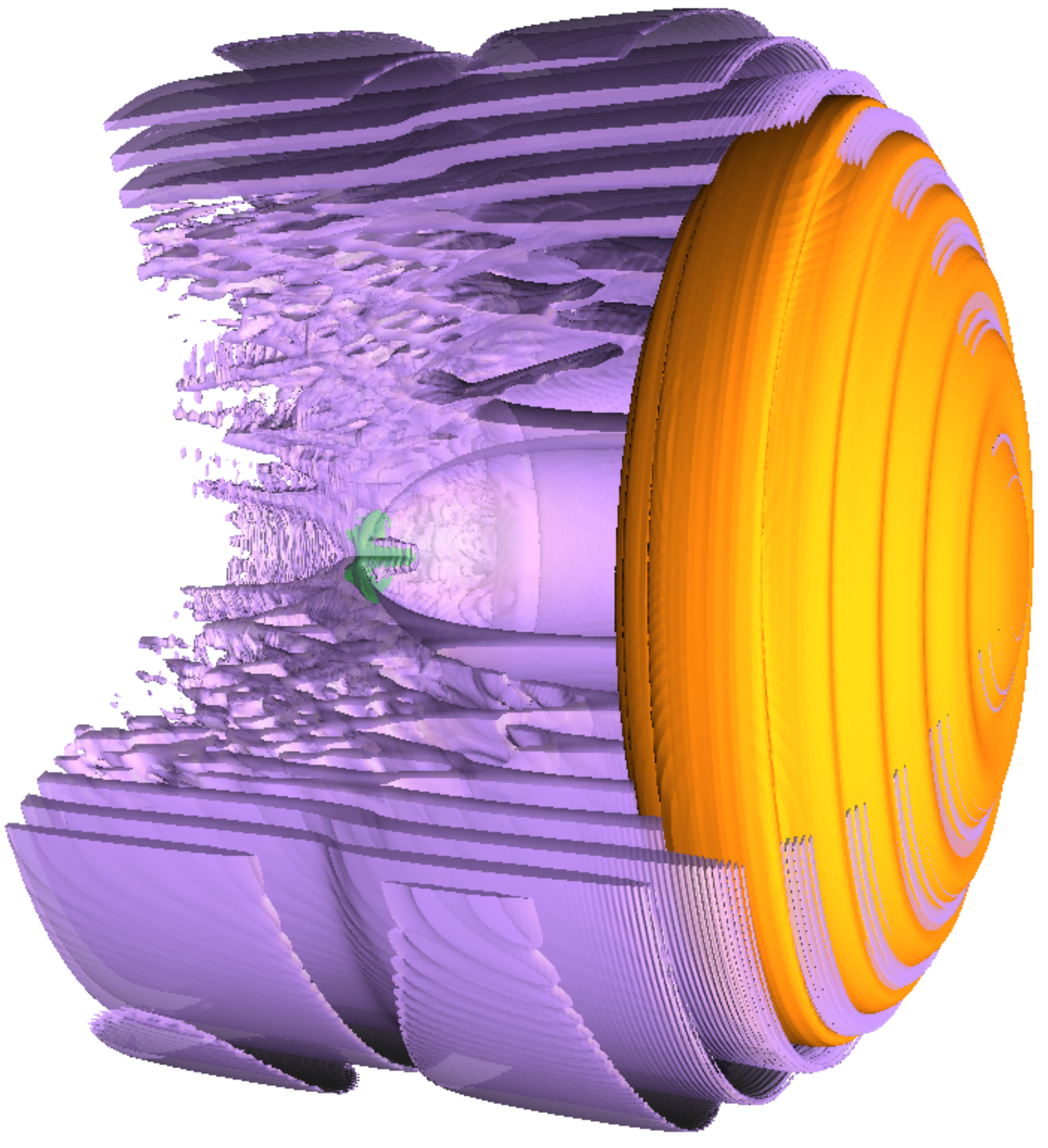}
        \put(78,5){(b)}
    \end{overpic}
    \caption{\label{fig:1} The illustration of the coaxial interference of the two laser pulses (a), which creates the onion-like plasma wake (b) in the proposed injection scheme. (a) The driver laser (red lines) is relatively loosely focused and the trigger laser (blue lines) is tightly focused, with the peak wavefront illustrated by solid lines and valley wavefront by dashed lines. Both lasers move to the right. The axial phase difference of the two lasers is $\Delta \phi_a$. The intersections of different types of lines, marked as black dots, are destructive points (rings in cylindrical geometry). The connective lines of the black dots become the sub-cavity sheath of the wakefield. The black arrows indicate the moving direction of the sheath. (b) The isosurface plot of the laser (orange), plasma (violet) and injected electron beam (cyan) from a 3D simulation. A quarter of the plasma is cut to show the interior.
}
\end{figure}

In this work, we propose a novel all-optical injection scheme that can generate high-quality electron beams with parameters beyond the above mentioned values, by utilizing the interference of two coaxial laser pulses. In this scheme, a driver laser pulse, containing the majority of the total energy, is relatively loosely focused with its Rayleigh range covering the injection region, while a trigger pulse, containing only a small portion of the total energy, is tightly focused shortly before the injection region, as illustrated in Fig.~\ref{fig:1} (a). Because of the different wavefront curvature of the two lasers, interference rings in the cross-section of the lasers are formed, pinching the plasma electrons by the traps of their ponderomotive force~\cite{PMoraPOP1997}. As a result, onion-like multiple sub-cavities are formed in the plasma wake as shown in Fig.~\ref{fig:1} (b). With the propagation of the two lasers, the innermost sub-cavity expands, which slows down the effective phase velocity of the wake. Similar to the density down-ramp injection, this process triggers injection of a beam with small slice energy spread and small emittance~\cite{AMOssaPRAB2017,XuPRAB2017}. The small slice energy spread is transformed to small energy spread of the entire beam by the self-dechirping effect, which is to be discussed later. The injection length is a few hundred micrometers, which is significantly longer than those in previous all-optical injection studies, allowing large amount of beam charge.

\begin{figure}
    \centering
    \begin{overpic}[width=0.45\textwidth]{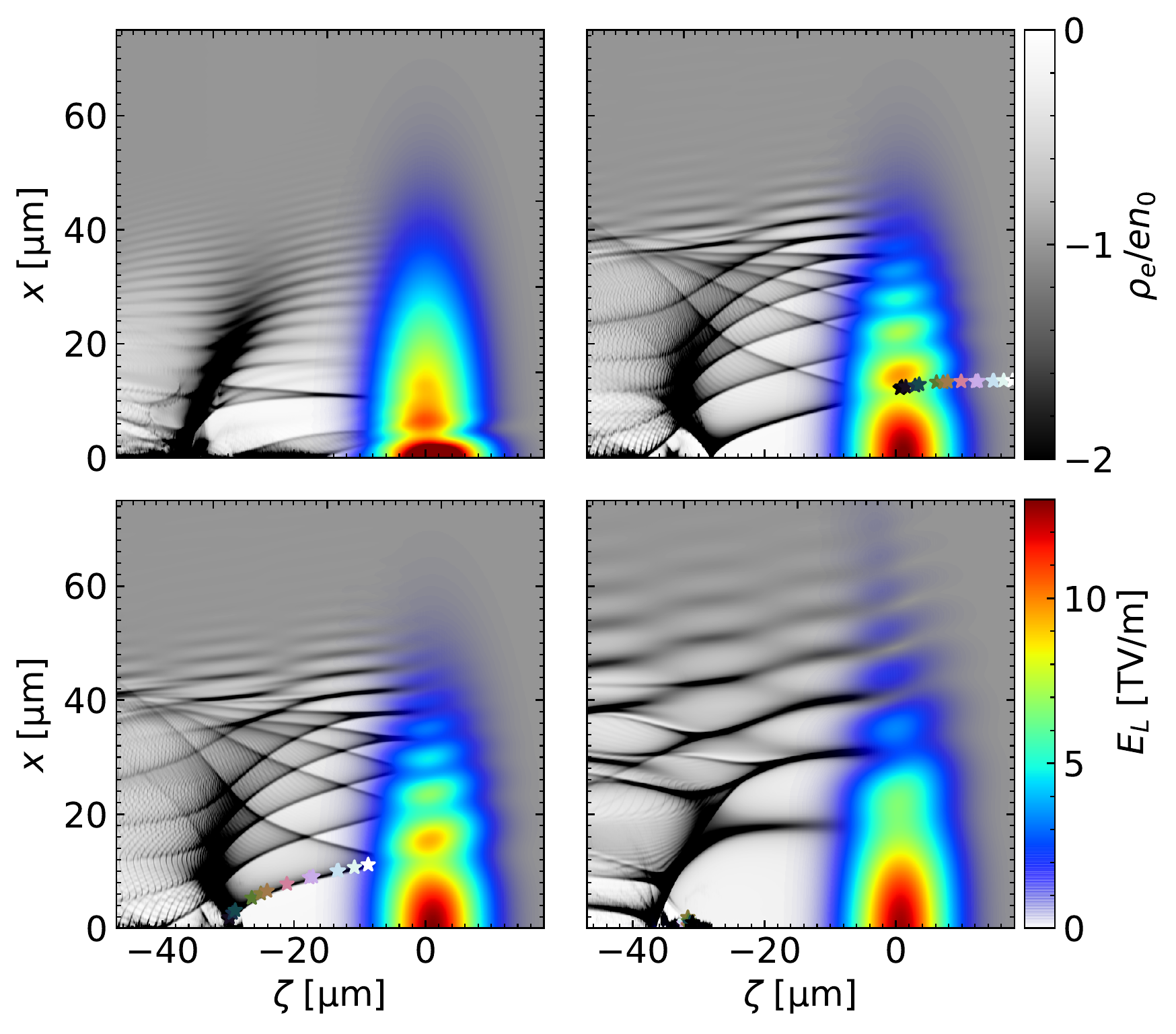}
        \put(11,81){\textcolor{white}{\bf(a) {\boldmath $t=0.21\ \rm ps$}}}
        \put(51,81){\textcolor{white}{\bf(b) {\boldmath $t=0.74\ \rm ps$}}}
        \put(11,41){\textcolor{white}{\bf(c) {\boldmath $t=0.84\ \rm ps$}}}
        \put(51,41){\textcolor{white}{\bf(d) {\boldmath $t=1.71\ \rm ps$}}}
    \end{overpic}
    \caption{\label{fig:PICsnap} A simulation of the proposed injection scheme. A driver laser with the peak power of 227.13 TW is focused to the waist of $w_{00}=30\ \rm \mu m$ and a trigger laser with the peak power of 1.88 TW is focused to the waist of $w_{01}=2\ \rm \mu m$. In the plots, the profile of the summation of the electric fields of both lasers is shown as $E_L$ by omitting the oscillation in the laser frequency scale. The snapshots show the innermost sub-cavity which (a) is not big enough to sustain an injection, (b) is big enough and the injection starts, (c) is expanding and the injection continues, (d) reaches the maximum size and the injection stops. A sample of injected electrons are marked as pentagrams.
}
\end{figure}

\begin{figure}
   \centering
    \begin{overpic}[width=0.45\textwidth]{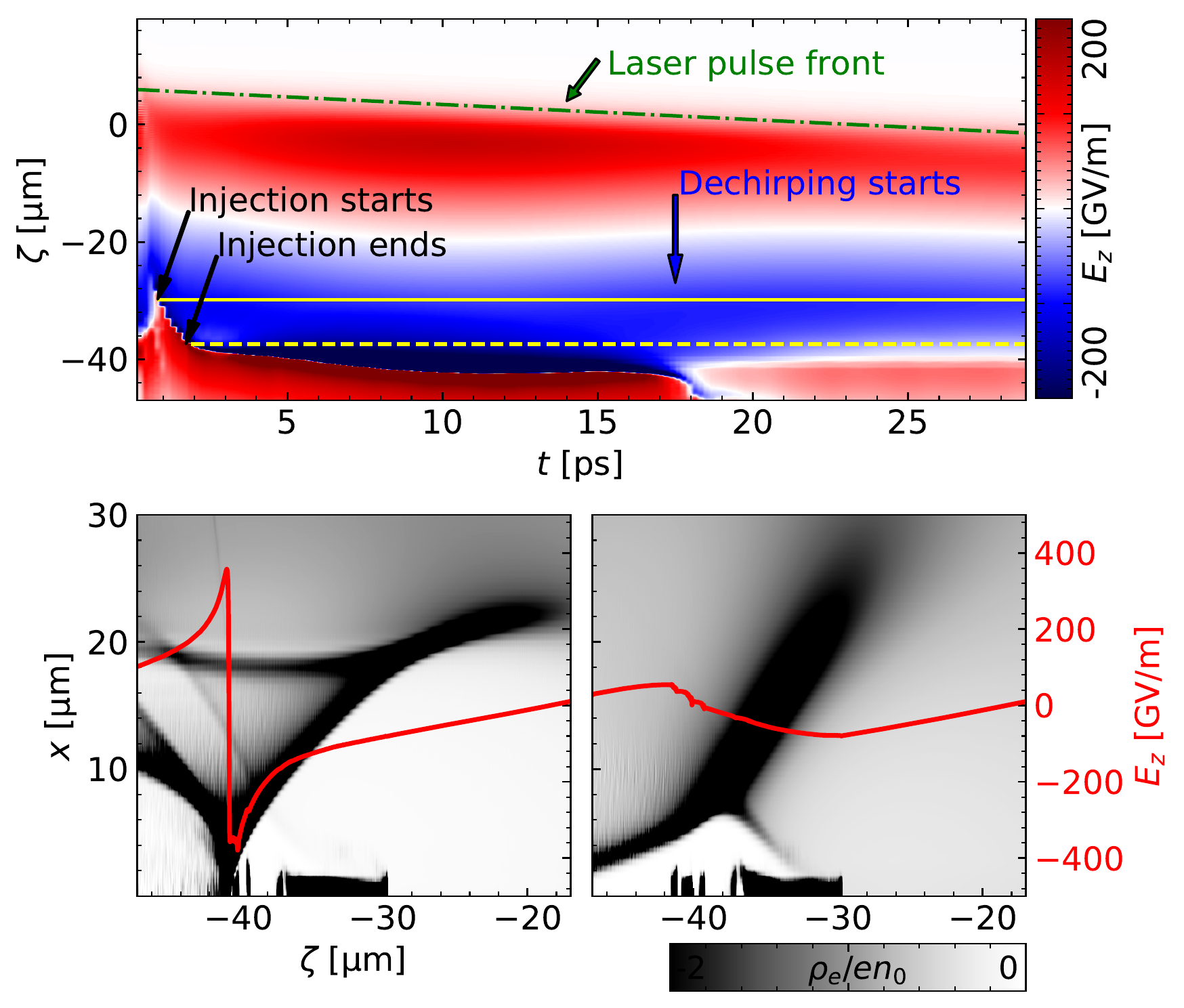}
        \put(81,80){(a)}
        \put(13,38){\textcolor{white}{\bf(b) {\boldmath $t=6.39\ \rm ps$}}}
        \put(52,38){\textcolor{white}{\bf(c) {\boldmath $t=28.06\ \rm ps$}}}
    \end{overpic}
    \begin{overpic}[width=0.44\textwidth]{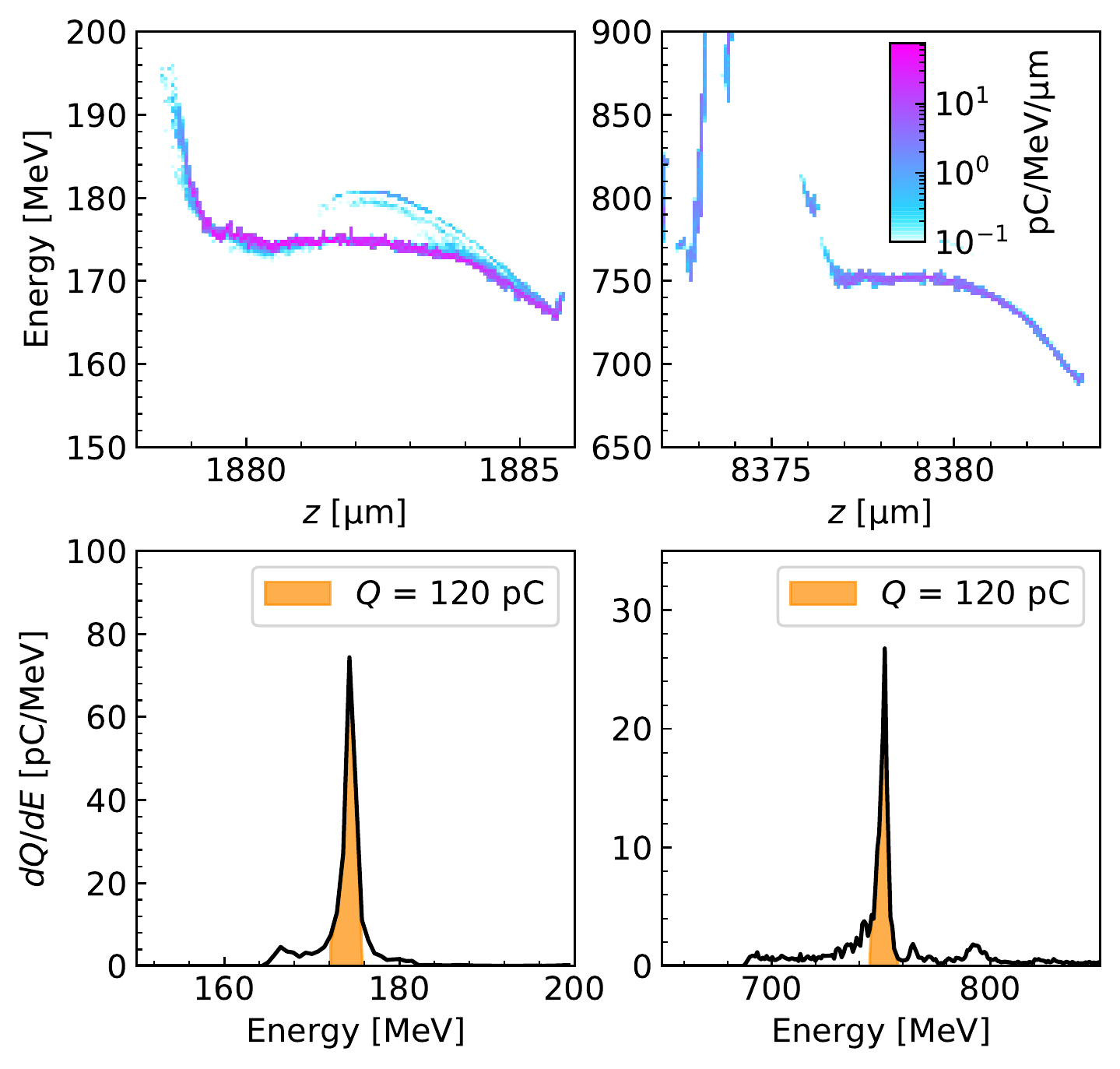}
        \put(14,59){(d) $t=6.39\ \rm ps$}
        \put(61,59){(e) $t=28.06\ \rm ps$}
        \put(14,12){(f)}
        \put(61,12){(g)}
    \end{overpic}
   \caption{\label{fig:dechirper} Plots showing the injection, self-dechirping and beam quality. (a) shows the evolution of the axial longitudinal electric field $E_z$. The contours of the laser pulse front, the head and tail of the injected electron beam are shown as green dash-dotted line, yellow solid and dashed lines, respectively. (b) and (c) show the snapshots at $t=6.39\ \rm ps$ and $28.06\ \rm ps$ (before and after self-dechirping). The phase spaces and energy spectra of the injected electron beam at these two instants of time are shown in (d) and (f), (e) and (g), respectively. The energy spread, the charge within threefold energy spread (colored areas), and the normalized emittance in the ($x$, $y$) directions are (f) 0.4\%, 120 pC, (0.15, 0.05) $\rm mm\cdot mrad$ and (g) 0.34\%, 120 pC, (0.25, 0.5) $\rm mm\cdot mrad$. The total charge of this beam is 170 pC which does not change with time after injection finishes.
}
\end{figure}
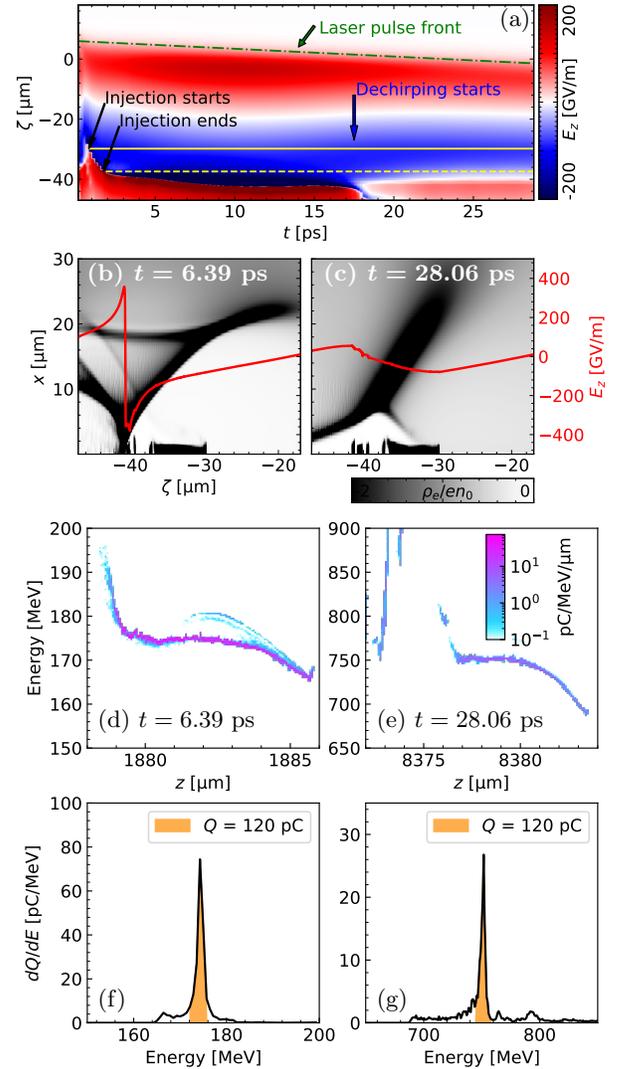

One example simulation using the code WarpX with spectral quasi-cylindrical formalism and PSATD solver~\cite{JLVayPOP2021} is shown in Fig.~\ref{fig:PICsnap}. In the plots, the longitudinal coordinate is transformed to the co-moving frame $\zeta=z-ct$, where $z$ is the longitudinal position with its zero point at the beginning of the flattop of the plasma density profile, $c$ is the speed of light in vacuum, and $t$ is the time with its zero point at the instant of time when the driver laser center reaches $z=0$. In this simulation, the unperturbed plasma density profile has a linear up-ramp from $z=-50\ \rm \mu m$ to 0 and a flattop from $z=0$ to $+\infty$, with the density of the flattop $n_0=1\times10^{18}\ \rm cm^{-3}$. The driver laser has a peak power of 227.13 TW and focused at $z=z_{f0}=450\ \rm \mu m$, with the focal waist of $w_{00}=30\ \rm \mu m$. The trigger laser has a peak power of 1.88 TW and focused at $z=z_{f1}=40\ \rm \mu m$, with the focal waist of $w_{01}=2\ \rm \mu m$. Both two lasers have the wavelength of $\lambda = 0.8\ \rm \mu m$, the pulse duration of $25\ \rm fs$ and are linearly polarized in the $y$ direction (perpendicular to the paper), but the trigger laser is $3.75\lambda$ ahead of the driver laser. The longitudinal size of the simulation box is $65\ \rm \mu m$ with 4096 cells, the transverse size is $500\ \rm \mu m$ with 1024 cells, the number of azimuthal modes is 2, and the time step interval is $dz/c$ where $dz$ is the longitudinal grid size. The number of macro particles per cell is 192 in the injection region, and 24 elsewhere. Due to the interference of the two lasers, rings with ponderomotive traps are formed, creating sub-cavities in the plasma wake. The expansion of the innermost sub-cavity sharply slows down the effective phase velocity of the wakefield from $t=0.6$ to $1.7\ \rm ps$ as shown in Fig.~\ref{fig:dechirper} (a), thus the injection of the background plasma electrons occurs, forming a beam with $\sim0.5\ \rm MeV$ slice energy spread. The longitudinal electric field exerts positive/negative chirping to the injected beam before/after $t=18\ \rm ps$, where the blowout regime is transformed to the partial-blowout regime, as shown in Fig.~\ref{fig:dechirper} (b) and (c), due to the pump-depletion effect of the driver pulse~\cite{LuWPRAB2007}. The energy spread has two minimums before and after the self-dechirping point as shown in Fig.~\ref{fig:dechirper} (d)-(g). One can see that a $170\ \rm pC$ electron beam with energy spread no more than $0.4\%$ and emittance no more than $0.5\ \rm mm\cdot mrad$ is generated.

A full three-dimensional (3D) PIC simulation with the same grid size and parameters as above has also been performed (only till the injection finishes due to limited computational resource). The injected beam has normalized emittance of $0.14\ \rm mm\cdot mrad$ in both transverse directions, charge of $150\ \rm pC$ and slice energy spread of $\sim 0.5\ \rm MeV$, which confirms the quasi-cylindrical simulation validity. This simulation at $t=1\ \rm ps$ is plotted as Fig.~\ref{fig:1} (b).

To explain the physics of the multi sheaths creation and injection, we write down the expression of the electric field of a Gaussian laser pulse in vacuum
\begin{equation}
    E(r,z,t)=E_0\frac{w_0}{w\left(z\right)}T\left(z-ct\right) \exp\left[-\frac{r^2}{w^2\left(z\right)}+i\phi\right], \label{eq:E}
\end{equation}
where $E_0$ is the peak electric field strength at focus, $w_0$ is the laser waist radius, $r=\sqrt{x^2+y^2}$ is the transverse position, $w\left(z\right)=w_0\sqrt{1+(z-z_f)^2/z_{R}^{2}}$ is the laser radius, $z_f$ is the focal position, $z_R=\pi w_0^2/\lambda$ is the Rayleigh length, $T\left(z-ct\right)=\exp\left[-{(z-ct-z_0)^2}/(c\tau)^2\right]$ is the temporal profile of the pulse, $z_0$ is the pulse center at $t=0$, $\tau$ is the pulse duration,
\begin{equation}
    \phi=-k\left[\frac{r^2}{2R\left(z\right)}+z-ct-z_0\right]+\phi_{\rm CEP}+\phi_{\rm G}
\end{equation}
is the phase, $k=2\pi/\lambda$ is the wave number, $\phi_{\rm CEP}$ is the carrier envelop phase,
\begin{equation}
    R(z)=z-z_f+\frac{z_R^2}{z-z_f}
\end{equation}
is the radius of the wavefront curvature, and
\begin{equation}
    \phi_{\rm G}=\arctan\left(\frac{z-z_f}{z_R}\right)
\end{equation}
is the Gouy phase. We neglect the plasma reaction to the laser pulses, because the interference region is short compared to the self-focusing length~\cite{MZengPOP2014}, and the phase velocity changes due to the plasma reaction are the same for the two lasers. We use an extra subscript 0 (1) for the driver (trigger) laser, the axial phase difference of the two lasers is
\begin{equation}
    \Delta\phi_a \equiv \left.\phi_1\right|_{r=0} - \left.\phi_0\right|_{r=0} = \Delta \phi_{-\infty}+\Delta\phi_{\rm G},
\end{equation}
where $\Delta \phi_{-\infty}=k\Delta z_0+\Delta\phi_{\rm CEP}$ is the phase difference far before focus, $\Delta z_0 = z_{01} - z_{00}$ is the difference of the centers of the two pulses, $\Delta\phi_{\rm CEP} = \phi_{\rm CEP1} - \phi_{\rm CEP0}$ is the difference of the carrier envelop phases, and $\Delta\phi_{\rm G} = \phi_{\rm G1} - \phi_{\rm G0}$. The phase difference for an arbitrary $r$ can be written as
\begin{equation}
    \begin{aligned}
        \Delta\phi = \phi_1 - \phi_0 &= \frac{kr^2}{2}\left(\frac{1}{R_0}-\frac{1}{R_1}\right)+\Delta\phi_a \\
                                 &\approx -\frac{kr^2}{2}\frac{1}{R_1}+\Delta\phi_a,
    \end{aligned}
\end{equation}
where only $z>z_{f1}$ (thus $R_1>0$) is considered, and the approximation is taken because $|R_0|\gg R_1$ in the injection region. The destructive inference rings have $\Delta\phi = -(2m+1)\pi$ where $m$ is an integer. These rings are ponderomotive traps, pinch the stream of the background electrons, and form sub-cavity sheaths. Thus the radii of the sub-cavities are
\begin{equation}
    \begin{aligned}
        r_c &= \sqrt{\frac{2R_1}{k}\left[(2m+1)\pi+\Delta\phi_a\right]}.
    \end{aligned}
\end{equation}
Without loosing of generality, we shift $\Delta\phi_a$ to the range $(-\pi,\pi]$ by adding a multiple of $2\pi$. Then $m$ should be a non-negative integer, and the radius of the innermost sub-cavity is
\begin{equation}
    r_c^* \equiv \left.r_c\right|_{m=0} = \sqrt{\frac{2R_1}{k}\left(\pi+\Delta\phi_a\right)}. \label{eq:rcstar}
\end{equation}
Based on simulation observations, we assume the longitudinal extent of the innermost sub-cavity can be estimated by $L_c \sim 2r_c^*$. Because the front half of the main wakefield bubble has deceleration electric field, and the sub-cavity cannot be longer than the main bubble, the necessary (but not sufficient) condition for injection is
\begin{equation}
    L_b/2 < L_c < L_b, \label{eq:Lc_range}
\end{equation}
where $L_b$ is the longitudinal extent of the main bubble. This implies $|z-z_{f1}|\gg z_{R1}$ and $|z-z_{f0}|\ll z_{R0}$ in the injection region, which lead to $\Delta \phi_G\approx \pi/2$ and $R_1\approx z-z_{f1}$. Thus Eq.~(\ref{eq:rcstar}) can be simplified in the injection region
\begin{equation}
    r_c^* \approx \sqrt{\frac{3\pi+2\Delta\phi_{-\infty}}{k}}\sqrt{z-z_{f1}},
\end{equation}
where $\phi_{-\infty}$ is shifted to the range $(-3\pi/2,\pi/2]$ by adding a multiple of $2\pi$. The increasing $r_c^*$ leads to an increasing $L_c$ and a slow-down of the wakefield phase velocity, which triggers injection of electrons from the sub-cavity sheath.

\begin{figure}
    \centering
    \begin{overpic}[width=0.45\textwidth]{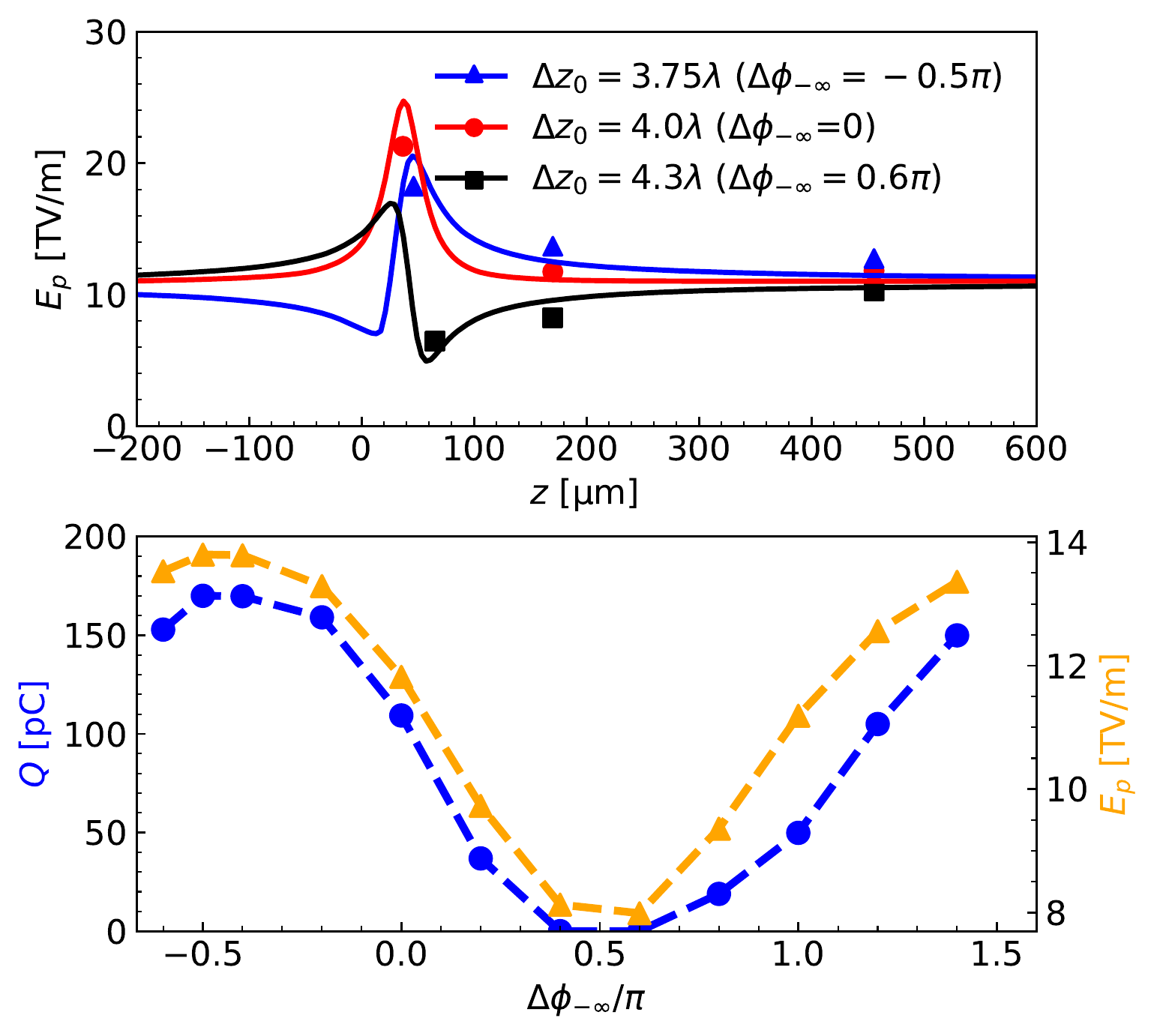}
        \put(14,57){(a)}
        \put(14,13){(b)}
    \end{overpic}
    \caption{\label{fig:Ep_vs_z} The dependence of the peak superimposed axial electric field of the two lasers $E_p$ and the injected charge to the initial phase difference of the two lasers $\Delta \phi_{-\infty}$, which is varied by changing $\Delta z_0$ while keeping $\Delta \phi_{\rm CEP}=0$. (a) $E_p$ vs.\ $z$ for different $\Delta \phi_{-\infty}$ cases. The solid lines represent the theoretical values and the scatter points are from simulations. (b) The beam charge after injection finishes (blue) and $E_p$ at $z=150\ \rm \mu m$ (yellow) vs.\ $\Delta \phi_{-\infty}$ obtained from simulations.
}
\end{figure}

To investigate the influence of the initial phase difference $\Delta \phi_{-\infty}$, we vary $\Delta z_0$ from $3.7\lambda$ to $4.7\lambda$ while keeping $\Delta \phi_{\rm CEP}=0$. The theoretical evolution of the peak axial electric field $E_p$, obtained by adding the electric field of the two lasers using Eq.~(\ref{eq:E}), is shown in Fig.~\ref{fig:Ep_vs_z} (a) as solid lines. The simulation results are plotted as scattered points, showing reasonably good agreement with the theory. The injected beam charge has strong correlation with $E_p$ at $z=150\ \rm \mu m$, which is a typical injection location, as shown in Fig.~\ref{fig:Ep_vs_z} (b). The injection is suppressed if $\Delta \phi_{-\infty}\approx \pi/2$ (thus $\Delta \phi_a\approx \pi$ with the same previous assumption $\Delta \phi_G\approx \pi/2$), because the interference is destructive on axis and the innermost sub-cavity does not have a clear sheath. It is worth noting that $\Delta z_0 = 3.75\lambda$ was used for the simulation shown in Figs.~\ref{fig:PICsnap} and \ref{fig:dechirper} which maximized the injection quantity.

In summary, we have proposed a new injection scheme in laser wakefield accelerators which generates electron beams with simultaneously small energy spread, small emittance and large amount of charge. In this scheme, one laser pulse is relatively loosely focused to drive the plasma wakefield, and another laser is tightly focused in the Rayleigh range of the former. The energy of the latter laser is only a small portion of the former, but the peak intensities are similar due to their focal spot size difference. Within a certain range, the tightly focused laser has convex wavefront, while the wavefront of the loosely focused laser is flat. Consequently, interference rings are formed, pinching the background electron stream and creating sub-cavities in the wakefield. Due to the fast varying wavefront curvature of the tightly focused laser, the innermost sub-cavities expands, slows down the effective phase velocity of the wakefield and triggers the injection of an electron beam with a few per mill energy spread and no more than $0.5\ \rm mm\cdot mrad$ emittance. The charge of the injected beam can be modulated by the initial phase difference of the two lasers, with the maximum exceeding $100\ \rm pC$ for a moderate total laser power $\sim 200\ \rm TW$.

\begin{acknowledgments}
This work is supported by Research Foundation of Institute of High Energy Physics, Chinese Academy of Sciences (Grant Nos.\ E05153U1, E15453U2, Y9545160U2 and Y9291305U2), and Key Research Program of Frontier Sciences of Chinese Academy of Sciences (Grant No. QYZDJ-SSW-SLH004).
\end{acknowledgments}

\bibliography{main}

\end{document}